# Structural, elastic, bonding, optoelectronic, and some thermo-physical properties of transition metal dichalcogenides ZrX$_2$ (X = S, Se, Te): Insights from ab-initio calculations

*Md. Mahamudujjaman, Md. Asif Afzal, R. S. Islam, S. H. Naqib**
*Department of Physics, University of Rajshahi, Rajshahi 6205, Bangladesh*
**Corresponding Author; Email: salehnaqib@yahoo.com*

**Abstract**

Transition metal dichalcogenides (TMDCs) belong to technologically important compounds. We have explored the structural, elastic, bonding, optoelectronic and some thermo-physical properties of ZrX$_2$ (X = S, Se, Te) TMDCs in details via ab-initio technique in this study. Elastic anisotropy indices, atomic bonding character, optoelectronic properties and thermo-physical parameters including melting temperature and minimum phonon thermal conductivity are investigated for the first time. All the TMDCs under investigation possess significant elastic anisotropy and layered structural features. ZrX$_2$ (X = S, Se, Te) compounds are fairly machinable, and ZrS$_2$ and ZrSe$_2$ are moderately hard. ZrTe$_2$, on the other hand, is significantly softer. Both covalent and ionic bondings contribute in the crystals. Electronic band structure calculations display semiconducting behavior for ZrS$_2$ and ZrSe$_2$ and metallic behavior for ZrTe$_2$. Energy dependent optoelectronic parameters exhibit good correspondence with the underlying electronic energy density of states features. ZrX$_2$ (X = S, Se, Te) compounds absorb ultraviolet radiation effectively. The reflectivity spectrum, R($\omega$), remains over 50% in the energy range from 0 eV to ~20 eV for ZrTe$_2$. Therefore, this TMDC has wide band and nonselective high reflectivity and can be used as an efficient reflector to reduce solar heating. Debye temperature, melting point and minimum phonon thermal conductivity of the compounds under study are low and show excellent correspondence with each other and also with the elastic and bonding characteristics.

## 1. Introduction

Transition metal dichalcogenides (TMDCs) [1–3] are crystalline solids with layered structure. The TMDCs have attracted notable attention of researchers primarily due to their potential electrochemical applications. They display widespread electronic properties such as metallic [4–6], semi-metallic [7–10], semiconducting [11–14], charge density waves [15,16], magnetic order [17–19] and superconductivity [20–22] attained mainly by electronic band structure engineering [23–25]. Binary TMDCs are materials that have unique mechanical, electric and optical properties, and have become potential rivals in the field of spintronics [26–30] as well as in optoelectronics [31–35]. I. Kar et al. [36] showed that electronic phase transition can occur from semiconducting to topological semi-metal in going from $ZrSe_2$ to $ZrTe_2$ by changing metal-chalcogen bond lengths. Binary TMDC topological semi-metals with nontrivial electronic structure exhibiting low-dissipation transport at room temperature can reduce power consumption and heating of conventional electronic equipments [37]. Furthermore, topological insulators and semi-metals have potential applications in quantum computation, novel optoelectronic and thermoelectric power harvesting devices.

One of the major advantages of binary TMDCs is tunable bandgap associated to a strong photoluminescence and large exciton binding energy that make them potential candidates for a variety of optoelectronic devices along with solar cells, photo detectors, light emitting diodes (LEDs), photo transistors, field effect transistors (FETs), and logic transistors [23,38–41]. As the TMDCs have layered structures and these layers are interconnected by weak interactions, foreign atoms or molecules can be inserted into the gap (the so called van der Waals gap) forming various intercalated compounds. Intercalation is an effective and efficient means for electronic band structure manipulation. Because of layered structures they can also be applied as dry lubricants [42,43].

In this study, three TMDC materials, $ZrX_2$ (X= S, Se, Te) are explored by the Kohn-Sham density functional theory (KS-DFT). There are some previous theoretical works on our chosen materials [36,44–46]. In those articles, structural, elastic, electronic and optical dielectric properties are studied. But still there are notable shortages of information on these materials. These lacks of information limit the feasibility of potential applications. Various thermo-physical properties, chemical bonding properties and charge density distribution of these materials have not been studied yet. Besides, a complete theoretical understanding of the optical properties is still lacking. So in this work, we have performed a broader investigation to bridge the existing knowledge gap. We have studied structural, elastic, bonding, charge density distribution, electronic, optical and thermo-physical properties of $ZrX_2$ (X = S, Se, Te). $ZrS_2$ and $ZrSe_2$ are semiconductors, and the band gaps fall in the visible-infrared region, that's why they could be potential candidates in photovoltaic device applications [47]. $ZrTe_2$, on the other hand, is metallic. The elastic, thermal and optical parameters exhibit a number of features pertinent to potential applications.

The rest of the paper is constructed in the following manner: In Section 2, we have discussed the computational methodology in detail. In Section 3, we have described the crystal structure, presented the computational results and analyzed those. Finally, the important findings of our calculations are discussed and summarized in Section 4.

## 2. Computational methodology

Density functional theory (DFT) is the most widely employed formalism for ab-initio calculations in crystalline solids where the ground state of crystalline system is found by solving the Kohn-Sham equation [48] with periodic boundary conditions (involving Bloch states). Here DFT based CAmbridge Serial Total Energy Package (CASTEP) code [49] has been used to explore various physical properties of the titled compounds. This code implements the total energy plane-wave pseudopotential method. In this study, we have used both local density approximation (LDA) and generalized gradient approximation (GGA) exchange-correlation functionals. It is well known that, LDA has a tendency of underestimating the lattice constants. On the other hand, GGA tends to overestimate the lattice constants. We have used both to check the suitability regarding the compounds of interest. In comparison with experimental lattice constants, GGA provides better results of ground state structural parameters for $ZrX_2$ (X = S, Se, Te). The most popular GGA functional, known as Perdew-Burke-Ernzerhof (PBE) scheme [50], tend to overestimate equilibrium volume more whereas it's modified version known as the PBEsol [51] accounts for better equilibrium volume. In PBEsol the two parameters of PBE are changed in order to satisfy the constraints that are more congenial for solids, except that, both functionals have the same analytical form. Therefore, the results reported herein are obtained using the GGA-PBEsol scheme. The electron-ion interaction is modeled by the Vanderbilt-type ultra-soft pseudopotential. This pseudopotential saves the computational time significantly without affecting the accuracy of the calculations to a large extent [52].

The valence electron configurations of Zr, S, Se, and Te have been taken as [$4d^2 5s^2$], [$3s^2 3p^4$], [$4s^2 4p^6$] and [$5s^2 5p^4$], respectively. Periodic boundary conditions are used to determine the total energies of each cell. In this paper, 18×18×12 k-points mesh is used based on Monkhorst-Pack scheme [53] for sampling the Brillouin zones of $ZrX_2$ (X = S, Se, Te) with a cut-off energy of 500 eV. On the other hand, to obtain a smooth Fermi surface of $ZrTe_2$, 32×32×18 k-points mesh has been used. The crystal geometry optimization is achieved through minimizing the total energy applying the Broyden-Fletcher-Goldferb-Shanno (BFGS) minimization technique [54]. The structure is relaxed up to a convergence threshold of 5×10$^{-6}$ eV-atom$^{-1}$ for energy, 0.01 eV Å$^{-1}$ for the maximum force, 0.02 GPa for maximum stress and 5×10$^{-4}$ Å for maximum displacement. The independent single crystal elastic constants $C_{ij}$, bulk modulus B, shear modulus G are calculated from the 'stress-strain' method contained within the CASTEP code. The electronic band structure, total density of states (TDOS) and partial density of states (PDOS) are obtained from the optimized geometry of $ZrX_2$ (X = S, Se, Te). The frequency-dependent optical properties are extracted from the estimated complex dielectric function, $\varepsilon(\omega) = \varepsilon_1(\omega) + i\varepsilon_2(\omega)$, which describes the frequency/energy dependent interactions of photons with electrons in solids. Using the Kramers-Kronig relationships, the real part $\varepsilon_1(\omega)$ of dielectric function $\varepsilon(\omega)$ has been obtained from the imaginary part $\varepsilon_2(\omega)$. The imaginary part, $\varepsilon_2(\omega)$, is calculated within the momentum representation of matrix elements of transition between occupied and unoccupied electronic states by employing the CASTEP supported formula expressed as:

$$\varepsilon_2(\omega) = \frac{2e^2\pi}{\Omega\varepsilon_0} \sum_{k,v,c} |\langle \psi_k^c | \hat{u} \cdot \vec{r} | \psi_k^v \rangle|^2 \, \delta(E_k^c - E_k^v - E) \tag{1}$$

where, $\Omega$ is the volume of the unit cell, $\omega$ is angular frequency of the incident electromagnetic wave (photon), $e$ is the electronic charge, $\psi_k^c$ and $\psi_k^v$ are the conduction and valence band wave functions at a given wave-vector $k$, respectively. The conservation of energy and momentum during the optical transition is implemented by the delta function. When the dielectric function $\varepsilon(\omega)$ is known, all the optical parameters such as refractive index, optical conductivity, reflectivity, absorption coefficient, and energy loss function can be computed from it. The Debye temperature and other thermo-physical parameters are calculated from elastic constants and elastic moduli of $ZrX_2$ (X = S, Se, Te) via the average sound velocity in the compounds of interest.

Many previous studies showed that, spin orbit coupling (SOC) has minimal effect on bulk physical properties, such as the optimized crystal structure, elastic constants, mechanical anisotropy, chemical bonding, thermo-physical behavior, bulk optical properties, etc [55–57]. Since our attention is on the bulk physical properties of the chosen materials which do not contain significant topological electronic features, we did not include SOC in the calculations to follow.

## 3. Results and analysis

### I. Structural properties

The $ZrX_2$ (X = S, Se, Te) compounds are isostructural and crystallize in *1T-CdI₂* type structure with space group *P-3m1* (no.164) in which octahedrally-coordinated X-Zr-X type sandwich layers are stacked with a periodicity of one layer possessing globally trigonal symmetry [58]. Due to van der Waals interaction, the interlayer bonding is weak while the intralayer Zr-X ionic bonding is strong [45]. The schematic crystal structure of $ZrX_2$ (X = S, Se, Te) is shown below in Fig. 1.

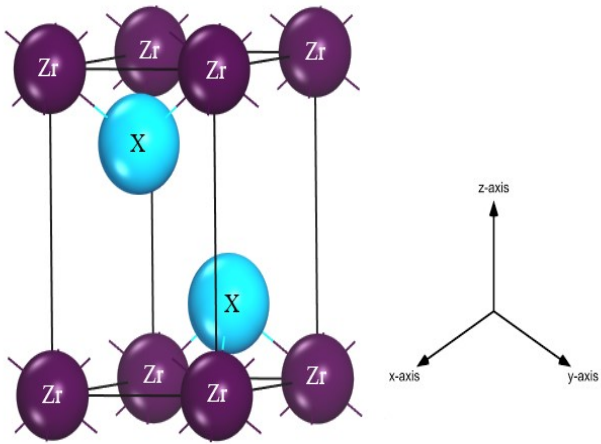

**Figure 1.** Schematic crystal structure of $ZrX_2$ (X = S, Se, Te) compounds.

Each unit cell of $ZrX_2$ (X = S, Se, Te) contains one formula unit and 10 atoms (8 Zr and 2 X) in total. Zr occupies 1c site (0 0 0) and X occupies 2d site (1/3 2/3 ±z). The only free structural parameter z, approximately equals to 1/4 in equilibrium. So in this study, we use 1/4 as the value of z. The optimized lattice parameters along with available theoretical and experimental lattice parameters are listed in the Table 1. The calculated lattice parameters of titled materials show good agreement with theoretical and experimental results. As mentioned earlier, application of LDA and GGA-PBE increases the deviations among theoretical and experimental structural parameters.

In Table 1 one observes that the equilibrium volumes obtained using GGA-PBEsol are slightly lower than that of experimental values which may appear somewhat unexpected. But it is worth noting that, the experimental structural parameters are determined at room temperature while the *ab-initio* calculations are performed assuming absolute zero temperature.

**Table 1.** Optimized lattice parameters (a = b in Å, c in Å, equilibrium volume $V_0$ in Å$^3$) of $ZrX_2$ (X = S, Se, Te) as compared to available theoretical and experimental data.

| Compound | a | c | c/a | $V_0$ | Ref. |
|---|---|---|---|---|---|
| ZrS$_2$ | 3.628 | 5.879 | 1.620 | 67.049 | This |
|  | 3.661 | 5.815 | 1.588 | 67.496 | [59]$^{Expt.}$ |
|  | 3.629 | 5.884 | 1.621 | 67.108 | [46]$^{Theo.}$ |
| ZrSe$_2$ | 3.713 | 6.060 | 1.632 | 72.378 | This |
|  | 3.766 | 6.150 | 1.633 | 75.538 | [36]$^{Expt.}$ |
|  | 3.732 | 6.134 | 1.643 | 73.987 | [46]$^{Theo.}$ |
| ZrTe$_2$ | 3.884 | 6.641 | 1.710 | 86.751 | This |
|  | 3.945 | 6.624 | 1.679 | 89.278 | [36]$^{Expt.}$ |
|  | 3.909 | 6.749 | 1.726 | 89.310 | [36]$^{Theo.}$ |

**II. Elastic properties**

In order to ensure structural/mechanical stability, it is very important to calculate elastic constants. Using single crystal elastic constants, various elastic properties can be determined. $ZrX_2$ (X = S, Se, Te) has trigonal crystal structure. This structure has six independent elastic constants ($C_{ij}$). These elastic constants are: $C_{11}$, $C_{33}$, $C_{44}$, $C_{12}$, $C_{13}$ and $C_{14}$. The calculated elastic constants of the titled compounds are listed in Table 2. According to Born-Huang conditions, a trigonal crystal system needs to satisfy the following criteria for mechanical stability [60]:

$$C_{11} - C_{12} > 0;$$

$$(C_{11} - C_{12})C_{44} - 2C_{14}^2 > 0;$$

$$(C_{11} + C_{12})C_{33} - 2C_{13}^2 > 0 \tag{2}$$

All the elastic constants of ZrX$_2$ (X = S, Se, Te) satisfy the mechanical stability criteria. This suggests that these TMDCs are mechanically stable. It is interesting to note that, small negative values of $C_{14}$ have no bearing on the mechanical stability of ZrS$_2$ and ZrSe$_2$. This may rather indicate that there are small internal strains in the optimized geometries of these crystals.

**Table 2.** Single crystal elastic constants, $C_{ij}$, of ZrX$_2$ (X = S, Se, Te) (all in GPa)

| Compound | $C_{11}$ | $C_{33}$ | $C_{44}$ | $C_{12}$ | $C_{13}$ | $C_{14}$ | Ref. |
|---|---|---|---|---|---|---|---|
| ZrS$_2$ | 130.67 | 22.64 | 6.86 | 23.07 | 0.83 | -3.08 | This |
|  | 131.47 | 29.93 | 52.92 | 25.63 | 4.25 | - | [45] |
| ZrSe$_2$ | 116.91 | 30.31 | 18.64 | 24.03 | 6.37 | -6.88 | This |
|  | 104.62 | 29.96 | 41.66 | 21.31 | 5.01 | - | [45] |
| ZrTe$_2$ | 67.65 | 32.30 | 6.94 | 12.95 | 8.07 | 1.07 | This |
|  | 69.00 | 26.00 | 31.00 | - | - | - | [44] |

Among the six independent single crystal elastic constants, $C_{11}$ and $C_{33}$ determine the ability of the crystal to resist the mechanical stress applied along the crystallographic a- (b-) and c-directions, respectively. From Table 2 we can see that, for all the compounds $C_{11}$ is greater than $C_{33}$. This suggests that the structure is more compressible in the c-direction than in a-direction. This is indicative of the layered nature of the compounds under study. This infers that the chemical bonding in the ab-plane is stronger than that in the out-of-plane direction. The resistance to shear deformation with respect to a tangential stress applied across the (100) plane in the [010] direction is quantified by elastic constant, $C_{44}$. Here $C_{44}$ < $C_{11}$ and $C_{33}$, which suggests that the compounds are more easily deformed by a shear in comparison to a unidirectional compression along any of the three crystallographic directions. The other elastic constants $C_{12}$, $C_{13}$, and $C_{14}$ are called off-diagonal shear components, which are related with compound's resistance due to various shape distortions.

Different types of elastic moduli, indicators and Poisson's ratio can be obtained from the single crystal elastic constants, $C_{ij}$. As single crystal samples are hard to synthesize and in many cases not practically applicable on large scale, information regarding elastic constants for polycrystalline aggregates are important for applications point of view. Table 3 exhibits the polycrystalline bulk modulus (B), Shear modulus (G), Pugh's ratio (B/G), Young's modulus (Y) and Poisson's ratio (η). In his treatment of elastic moduli for polycrystalline solids, Voigt [61]assumed that strain is uniform throughout the aggregate and derived effective isotropic stiffnesses in terms of space averages of the single crystal elastic stiffnesses over all possible orientations. This approximation gives us the upper limit of the polycrystalline elastic moduli. Reuss [62], on the other hand, assumed uniform stress throughout the polycrystal and derived effective isotropic elastic compliances in terms of single-crystal elastic compliances averaged over all orientations resulting in the lower limit of the polycrystalline elastic moduli. The real value lies between Voigt and Reuss bounds. Hill [63] later proposed the arithmetic average of the two limits, which closely represents the practical situation.

The bulk modulus describes the material's response to the volume changing hydrostatic pressure and the shear modulus measures the material's resistance against shape deforming shear stress. From Table 3, it is evident that B is greater than G for all the compounds. Therefore, the mechanical stabilitiesof$ZrX_2$ (X = S, Se, Te) are expected to be controlled by shearing strain.

**Table 3.** Elastic moduli (all in GPa), Pugh's ratio, and Poisson's ratio of $ZrX_2$ (X = S, Se, Te) compounds.

| Compound | $B_V$ | $B_R$ | $B_H$ | $G_V$ | $G_R$ | $G_H$ | B/G | Y | η | Ref. |
|---|---|---|---|---|---|---|---|---|---|---|
| ZrS$_2$ | 37.052 | 17.782 | 27.417 | 30.786 | 12.459 | 21.622 | 1.267 | 51.365 | 0.187 | This |
| | 40.120 | 23.330 | 31.730 | 31.490 | 16.210 | 23.850 | 1.330 | 57.22 | 0.200 | [45] |
| ZrSe$_2$ | 37.523 | 23.805 | 30.664 | 31.901 | 23.498 | 27.699 | 1.107 | 63.867 | 0.152 | This |
| | 33.540 | 22.450 | 27.990 | 26.560 | 17.470 | 22.020 | 1.270 | 52.33 | 0.190 | [45] |
| ZrTe$_2$ | 25.090 | 21.905 | 23.497 | 17.478 | 11.781 | 14.629 | 1.606 | 36.346 | 0.242 | This |
| | - | - | - | - | - | - | - | - | - | - |

The ratio B/G, called the Pugh's ratio is an indicator of the ductile or brittle nature of a compound. If the value of Pugh's ratio is greater than 1.75, the material is predicted to be ductile in nature; otherwise the material should exhibit brittleness [64]. Thus all the materials in this work are expected to show brittle behavior. Poisson's ratio (η) is a measure of materials deformation (expansion or contraction) along the perpendicular direction of loading. It also gives a measure of the stability of solids against shear. If η = 0.5, no volume change occur during elastic deformation. For $ZrX_2$ (X = S, Se, Te) compounds, η is much lower than 0.5 which means a large volume change is associated during elastic deformation. Poisson's ratio is an important parameter that determines various mechanical properties of crystalline solids. It can predict the ductility or brittleness of materials with the critical value of 0.26. If η is less (greater) than 0.26, the material is brittle (ductile). Thus, this judgment criteria tells us that $ZrX_2$ (X = S, Se, Te) compounds are brittle. This is consistent with the result obtained from the Pugh's ratio. The value of η also can predict the nature of interatomic forces in solids [65,66]. If η remains between 0.25 and 0.50, central force interaction will dominate. Otherwise non-central force will dominate. Thus, in $ZrX_2$ (X = S, Se, Te) non-central force should dominate the atomic bonding. In completely ionic bonded solids, the value of η is approximately 0.33, while in purely covalent bonded crystal, it is around 0.10. This implies that, in ZrS$_2$ and ZrSe$_2$ covalent bondings should be dominant and in ZrTe$_2$ ionic bonding should be dominant. The overall bonding characters of the TMDCs under consideration should exhibit mixed character with different proportions of covalent and ionic contributions. Besides, Poisson's ratio signifies the level of plasticity of a solid against shear. Larger the Poisson's ratio better is the plasticity.

A number of useful mechanical performance indicators, namely the machinability index ($\mu^M$), Kleinman parameter ($\zeta$) and Vickers hardness ($H_v$) are calculated using the following widely employed equations [67,68] and are listed in Table 4.

$$\mu^M = \frac{B}{C_{44}} \tag{3}$$

$$\zeta = \frac{C_{11} + 8C_{12}}{7C_{11} + 2C_{12}} \tag{4}$$

$$H_V = \frac{(1-2\eta)Y}{6(1+\eta)} \tag{5}$$

The machinability index is also a measure of dry lubricating nature of a material. A high value of $\mu^M$ indicates excellent lubricating properties, i.e., lower friction. The value of $\mu^M$ in $ZrX_2$ (X = S, Se, Te) implies a good level of machinability which compares favorably to many technologically important MAX phase nanolaminates [69–73]. The machinability indices of $ZrS_2$ and $ZrTe_2$ are particularly high. Generally, the value of Kleinman parameter lies between 0 and 1. According to Kleinman, the lower limit of $\zeta$ represents significant contribution of bond stretching or contracting to resist external stress whereas the upper limit corresponds to significant contribution of bond bending to resist external load. From Table 4, it is evident that, mechanical strength in $ZrX_2$ (X = S, Se, Te) is mainly derived from the bond stretching/contracting contribution. The hardness of a solid is essential to understand elastic and plastic properties.

The bulk modulus along a-, b- and c-axis (known as directional bulk modulus) and isotropic bulk modulus ($B_{relax}$) are defined as follows [68] and are also enlisted in Table 4.

$$B_a = a\frac{dP}{da} = \frac{\Lambda}{1+\alpha+\beta} \tag{6}$$

$$B_b = b\frac{dP}{db} = \frac{B_a}{\alpha} \tag{7}$$

$$B_c = c\frac{dP}{dc} = \frac{B_a}{\beta} \tag{8}$$

$$B_{relax} = \frac{\Lambda}{(1+\alpha+\beta)^2} \tag{9}$$

where, $\Lambda = C_{11} + 2C_{12}\alpha + C_{22}\alpha^2 + 2C_{13}\beta + C_{33}\beta^2 + 2C_{23}\alpha\beta$

and

$$\alpha = \frac{\{(C_{11} - C_{12})(C_{33} - C_{13})\} - \{(C_{23} - C_{13})(C_{11} - C_{13})\}}{\{(C_{33} - C_{13})(C_{22} - C_{12})\} - \{(C_{13} - C_{23})(C_{12} - C_{23})\}}$$

$$\beta = \frac{\{(C_{22} - C_{12})(C_{11} - C_{13})\} - \{(C_{11} - C_{12})(C_{23} - C_{12})\}}{\{(C_{22} - C_{12})(C_{33} - C_{13})\} - \{(C_{12} - C_{23})(C_{13} - C_{23})\}}$$

The calculated $B_{relax}$ using Eqn. 9 gives the same value as the Reuss approximation does. α and β are the relative change of the b- and c- axis as a function of the deformation of the a axis. The small value of $B_c$ compared to $B_a$ and $B_b$ indicates that $ZrX_2$ (X = S, Se, Te) are more compressible when stress is applied along c-direction than along a- or b-direction.

**Table 4.** The machinability index ($\mu^M$), Kleinman parameter (ζ), Vickers hardness ($H_v$ in GPa), bulk modulus ($B_{relax}$ in GPa), bulk modulus along a-, b- and c-axis ($B_a$, $B_b$, $B_c$ in GPa), α and β of $ZrX_2$ (X = S, Se, Te) compounds.

| Compound | $\mu^M$ | ζ | $H_v$ | $B_{relax}$ | $B_a$ | $B_b$ | $B_c$ | α | β | Ref. |
|---|---|---|---|---|---|---|---|---|---|---|
| $ZrS_2$ | 3.996 | 0.328 | 6.582 | 17.782 | 102.66 | 102.66 | 14.723 | 1 | 6.972 | This |
| $ZrSe_2$ | 1.644 | 0.356 | 6.411 | 23.805 | 175.091 | 175.091 | 32.695 | 1 | 5.355 | This |
| $ZrTe_2$ | 3.385 | 0.342 | 2.513 | 21.905 | 102.083 | 102.083 | 38.375 | 1 | 2.66 | This |

It is instructive to notice that $ZrS_2$ and $ZrSe_2$ are reasonably hard compounds while $ZrTe_2$ is significantly softer. All the results displayed in Table 4 are novel and there is no existing data to be compared.

It is worth stating that elastic anisotropy has influence on development of plastic deformation in crystals, formation and propagation of microscale cracks in ceramics, plastic relaxation in the thin films etc. Therefore, it is quite important to evaluate the elastic anisotropy factors for solids to predict their behavior under different conditions of external stresses [73–75]. The elastic/mechanical anisotropy indices of $ZrX_2$ (X = S, Se, Te) are investigated in this section. We have calculated Zener anisotropy factor (A), shear anisotropy factors ($A_1$, $A_2$, $A_3$), percentage anisotropy in compressibility ($A_B$) and shear ($A_G$), universal anisotropy factor ($A^U$, $d_E$), Equivalent Zener anisotropy factor ($A^{eq}$) and disclosed those in Table 5. The following widely used relations are used to calculate anisotropy factors [76–78]:

$$A = \frac{2C_{44}}{C_{11} - C_{12}} \quad (10)$$

$$A_1 = \frac{4C_{44}}{C_{11} + C_{33} - 2C_{13}} \quad (11)$$

$$A_2 = \frac{4C_{55}}{C_{22} + C_{33} - 2C_{23}} \quad (12)$$

$$A_3 = \frac{4C_{66}}{C_{11} + C_{22} - 2C_{12}} \tag{13}$$

$$A_B = \frac{B_V - B_R}{B_V + B_R} \tag{14}$$

$$A_G = \frac{G_V - G_R}{G_V + G_R} \tag{15}$$

$$A^U = \frac{B_V}{B_R} + 5\frac{G_V}{G_R} - 6 \geq 0 \tag{16}$$

$$d_E = \sqrt{A^U + 6} \tag{17}$$

$$A^{eq} = \left(1 + \frac{5}{12}A^U\right) + \sqrt{\left(1 + \frac{5}{12}A^U\right)^2 - 1} \tag{18}$$

The calculated values of A, $A_1$ and $A_2$ are significantly different from 1. These anisotropy factors imply that $ZrX_2$ (X = S, Se, Te) are strongly anisotropic with respect to shearing stress along different crystal planes. But there is an exception, the shear anisotropy factor $A_3$ = 1 for $ZrX_2$ (X = S, Se, Te), which means that for {001} shear planes between ⟨110⟩ and ⟨010⟩ directions, the compounds are isotropic in nature. For $A_B$ and $A_G$, zero value represents elastic isotropy and a value of 1 corresponds to highest anisotropy. Therefore, from Table 5, it is evident that the compounds under study are anisotropic in compressibility and shear. An anisotropy factor $A^U$, known as the universal anisotropy factor because of its applicability to all kinds of crystal symmetries was defined by Ranganathan and Ostoja-Starzewski [79]. For elastically isotropic crystals $A^U$ = 0, while any other positive value indicates anisotropy. The calculated values of $A^U$ suggests that, $ZrX_2$ (X = S, Se, Te) compounds possess large amount of anisotropy in elastic properties. For $A^{eq}$, a value of 1 represents isotropy, while any other value indicates anisotropy. Thus, the values of $A^{eq}$ for $ZrX_2$ (X = S, Se, Te) indicate that the compounds under investigations are highly anisotropic. All the anisotropy indices reflect of structural anisotropy which originates from the anisotropy in the bonding strengths in different directions in the unit cell of the crystal.

The universal log-Euclidean index is defined by the following equation [77]:

$$A^L = \sqrt{\left[\ln\left(\frac{B_V}{B_R}\right)\right]^2 + 5\left[\ln\left(\frac{C_{44}^V}{C_{44}^R}\right)\right]^2} \tag{19}$$

where,

$$C_{44}^R = \frac{5}{3}\frac{C_{44}(C_{11}-C_{12})}{3(C_{11}-C_{12})+4C_{44}} \quad \text{is the Reuss value of } C_{44}$$

$$C_{44}^V = C_{44}^R + \frac{3}{5}\frac{(C_{11}-C_{12}-2C_{44})^2}{3(C_{11}-C_{12})+4C_{44}} \quad \text{is the Voigt value of } C_{44}$$

According to Kube and Jong [80], the value of $A^L$ lies in the range $0 \leq A^L \leq 10.26$ for inorganic crystalline compounds and 90% of these compounds have $A^L < 1$. For perfect isotropy, $A^L = 0$. Based on the value of $A^L$, it is difficult to ascertain whether a material is layered/lamellar or not. But, the majority (78%) of these inorganic crystalline compounds with high $A^L$ value exhibit layered/lamellar structure [80]. Based on the above discussion, we predict that the compounds under investigation are highly anisotropic and layered in nature. The compound ZrS$_2$ is expected to exhibit very high level of layered feature in particular.

**Table 5.** Elastic anisotropy indices of ZrX$_2$ (X = S, Se, Te) compounds.

| Compound | A | $A_1$ | $A_2$ | $A_3$ | $A_B$ | $A_G$ | $A^U$ | $d_E$ | $A^{eq}$ | $A^L$ | $A_{B_a}$ | $A_{B_c}$ | Ref. |
|---|---|---|---|---|---|---|---|---|---|---|---|---|---|
| ZrS$_2$ | 0.127 | 0.18 | 0.18 | 1 | 0.351 | 0.423 | 8.438 | 3.799 | 8.92 | 3.763 | 1 | 0.143 | This |
| ZrSe$_2$ | 0.401 | 0.554 | 0.554 | 1 | 0.223 | 0.151 | 2.364 | 2.892 | 3.699 | 1.199 | 1 | 0.186 | This |
| ZrTe$_2$ | 0.253 | 0.331 | 0.331 | 1 | 0.067 | 0.194 | 2.563 | 2.926 | 3.878 | 2.123 | 1 | 0.375 | This |

The anisotropies of the bulk modulus along a- and c-axis are defined as [68]:

$$A_{B_a} = \frac{B_a}{B_b} = \alpha \tag{20}$$

$$A_{B_c} = \frac{B_c}{B_b} = \frac{\alpha}{\beta} \tag{21}$$

The calculated values of $A_{B_a}$ and $A_{B_c}$ are also enlisted in Table 5 and the values represent that the bulk modulus for ZrX$_2$ (X = S, Se, Te) along a-axis is isotropic and along c-axis is anisotropic.

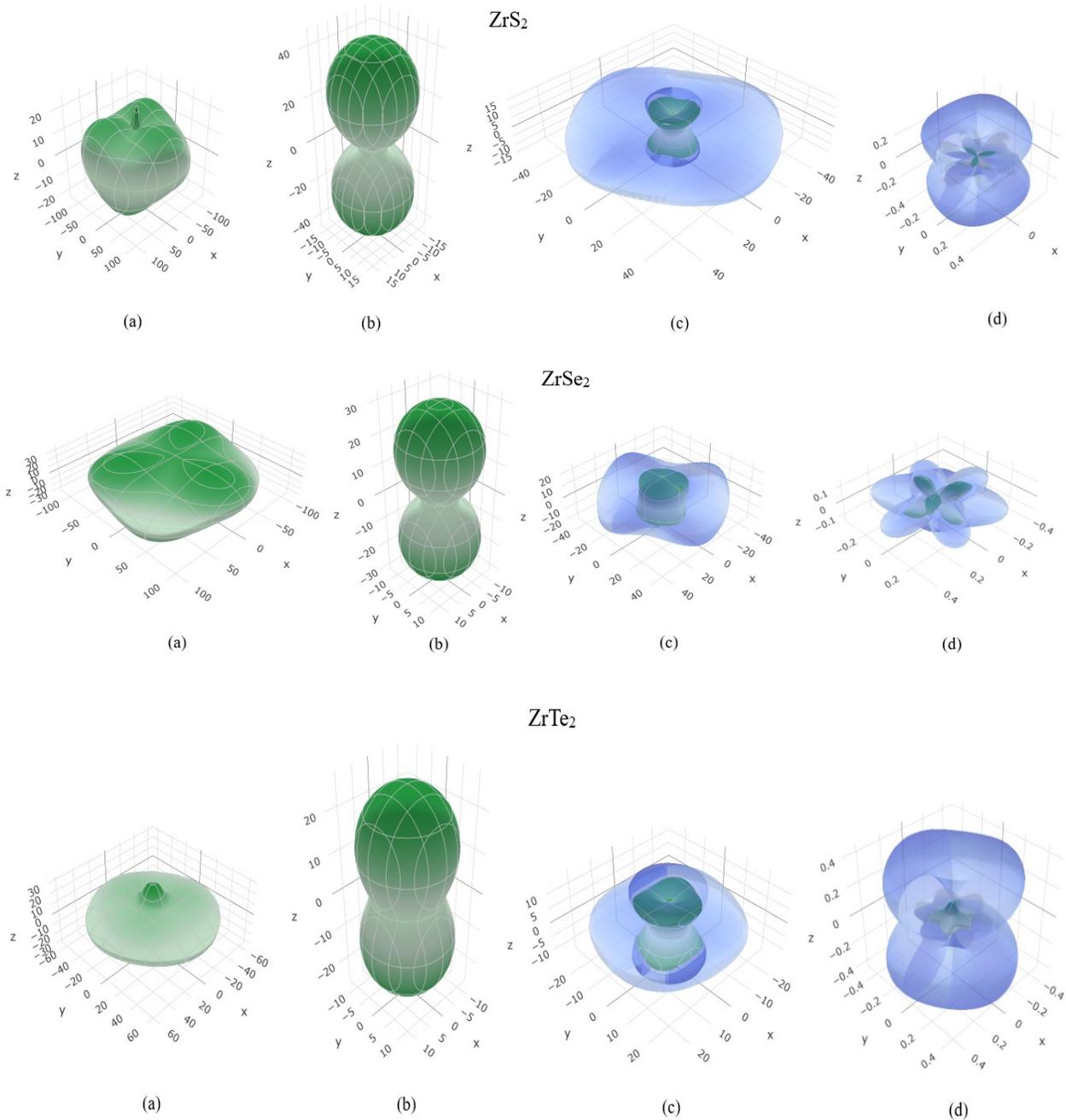

**Figure 2.** 3D directional dependences of (a) Young modulus (b) shear modulus (c) linear compressibility and (d) Poisson's ratio for $ZrX_2$ (X = S, Se, Te) compounds.

For isotropic solids, three dimensional (3D) direction dependent Young modulus, shear modulus, linear compressibility and Poisson's ratio should exhibit spherical shapes, while any deviation from spherical shape would indicate anisotropy. We have shown ELATE [81] generated 3D plots of directional dependence of Young modulus, shear modulus, linear compressibility and Poisson's ratio for $ZrX_2$ (X = S, Se, Te) compounds in Fig. 2 above.

## III. Electronic band structure and density of states

### (a) Electronic band structure

Electronic band structures for optimized crystal structure of ZrX$_2$ (X = S, Se, Te) along several high symmetry directions in the first Brillouin zone (BZ) are depicted in Fig. 3. The band structures of ZrS$_2$ and ZrSe$_2$ show that they have indirect minimal band gaps between valence band maxima at $\Gamma(k = (0,0,0))$ and conduction band minima at $L(k = (1/2,0,1/2))$. The direct and indirect band gap values of ZrS$_2$ and ZrSe$_2$ along with available experimental and theoretical values are given in Table 6.

**Table 6.** Calculated band gaps in eV of ZrS$_2$ and ZrSe$_2$ with available theoretical and experimental values.

| Compounds | Band gap ($E_g$) type | Theoretical $E_g$ [Ref.] | Experimental $E_g$ [Ref.] |
|---|---|---|---|
| ZrS$_2$ | Indirect | 0.74 [This] | 1.70 [82] |
| | | 0.79 [46] | |
| | Direct | 1.50 [This] | 2.10 [82] |
| | | 1.45 [46] | |
| ZrSe$_2$ | Indirect | 0.14 [This] | 1.18 [82] |
| | | 0.21 [46] | |
| | Direct | 1.04 [This] | 1.61 [82] |
| | | 0.89 [46] | |

It is well known that, in DFT use of GGA and LDA type exchange-correlations functionals underestimate the band gaps of semiconductors and insulators [82]. We have displayed the GGA results in Table 3 since it optimized the crystal structures most effectively. Though, indirect band gaps deviate from the other theoretical and available experimental values, direct band gaps show better agreement with theoretical [46] and experimental results [82]. The band structure diagram of ZrTe$_2$ shows that a number of bands (red and green curves) cross the Fermi level ($E_F$), which is indicated by a horizontal line placed at 0 eV. Therefore, ZrTe$_2$ is a metal. Highly dispersive nature of the bands crossing the Fermi level near $\Gamma$-point for ZrTe$_2$ is indicative of high mobility of the charge carriers in this compound. It is interesting to note that the band crossing near the $\Gamma$-point displays hole-like feature while the crossings near the A- and L-point of the BZ exhibit electron-like dispersions. Overall, for all three compounds under investigation, the curves along $\Gamma$-A, H-K directions are less dispersive, which indicates high effective mass of charge carriers and consequently low mobility in these directions. On the other hand, the band curves along A-H and K-$\Gamma$ directions show high dispersion, which indicates low effective mass and high mobility of electrons. Thus, anisotropy in charge transport within and out of ab-plane should be observed.

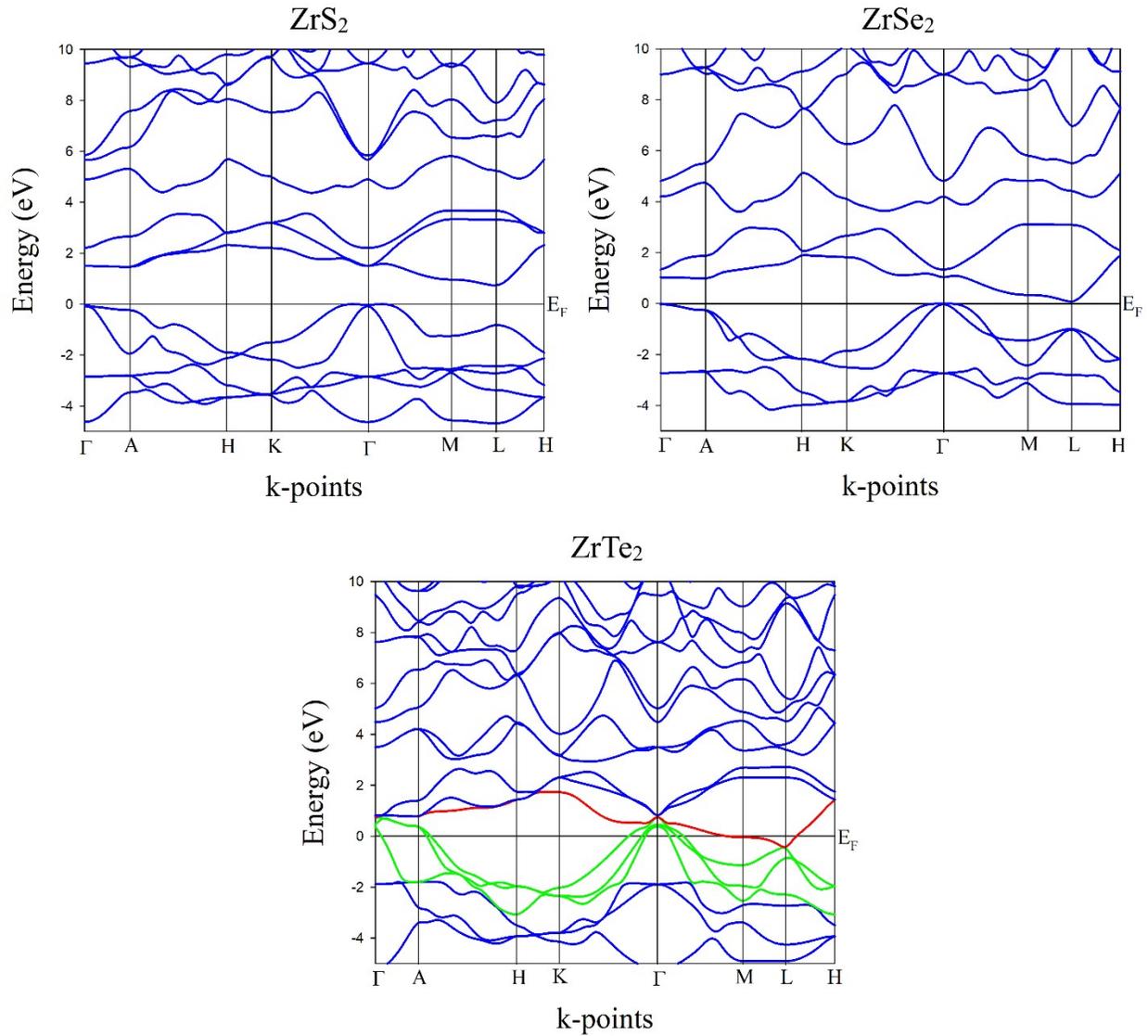

**Figure 3.** The band structure of ZrX$_2$ (X = S, Se, Te) compounds along the high symmetry directions of the k-space within the first Brillouin zone.

**(b) Electronic energy density of states**

To comprehend the electronic properties of ZrX$_2$ (X = S, Se, Te) in greater depth, the electronic energy density of states in the valence and conduction bands have been calculated using the energy dispersion curves shown in Fig. 3. Figure 4 shows the calculated total density of states (TDOS) and atom resolved partial density of states (PDOS) of ZrX$_2$ (X = S, Se, Te) compounds as a function of energy, (E - E$_F$). The vertical dashed line at 0 eV represents the Fermi level, E$_F$. The TDOS of ZrS$_2$ and ZrSe$_2$ at Fermi level is zero, which indicates the semiconducting nature of these materials, whereas the TDOS of ZrTe$_2$ is ~1.0 electronic states/eV indicating its metallic nature. This value for ZrTe$_2$ is in fair agreement with the previous work with a value of 0.70 electronic states/eV [83].

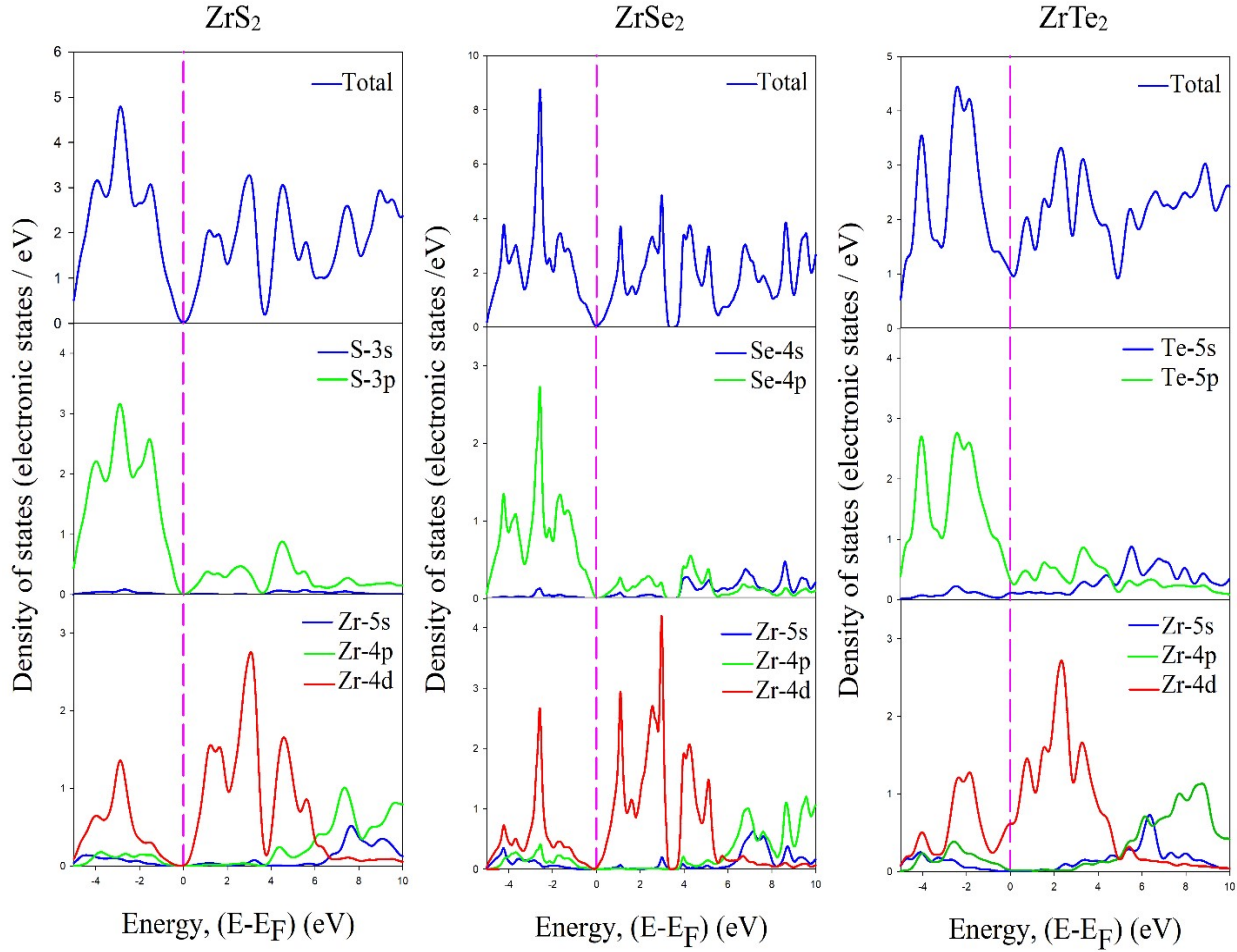

**Figure 4.** Total and partial density of states of ZrX$_2$ (X = S, Se, Te) compounds.

The valence bands of ZrS$_2$ and ZrSe$_2$ mainly consist of S-3p, Zr-4d electronic states and Se-4p, Zr-4d electronic states, respectively with almost equal contribution. The contributions of S-3s, Se-4s, Zr-5s, and Zr-4p orbitals to the valence band are minimal. In ZrTe$_2$, the band mainly consists of Te-5p and Zr-4d electronic states having the highest contribution from Te-4p. Both these bands due to Te-5p and Zr-4d electrons cross the Fermi level and give rise to metallic behavior. On the other hand, in the conduction bands of these compounds near the Fermi level dominant contributions come from the Zr-4d electronic states. There is significant overlap in energy between the S-3p and Zr-4d bands in ZrS$_2$, Se-4p and Zr-4d bands in ZrSe$_2$, and Te-5p and Zr-4d bands in ZrTe$_2$ compounds. Such overlaps are suggestive of covalent bonding between the electronic states involved. The nearest peak at the negative energy below the Fermi level in TDOS is known as bonding peak, while the nearest peak at the positive energy is the anti-bonding peak. The energy gap between these peaks is called the pseudogap. A pseudogap very close to $E_F$ is an indication of high structural stability [84–87]. In ZrX$_2$ (X = S, Se, Te) compounds bonding and anti-bonding peaks are within 2 eV from Fermi level. The close proximity of the peaks in the TDOS to the Fermi energy indicates that it might be possible to change the electronic phase of these compounds by suitable atomic substitution (alloying) or by applying pressure.

## IV. Charge density distribution and bonding properties

### (a) Charge density distribution

To study the bonding nature between the atoms of ZrX$_2$ (X = S, Se, Te), the valence electron charge density distributions within the (100) and (001) planes are depicted in Figure 5. The color scale between the maps represents the local electron density. Red and blue colors indicate lowest and highest electron density, respectively.

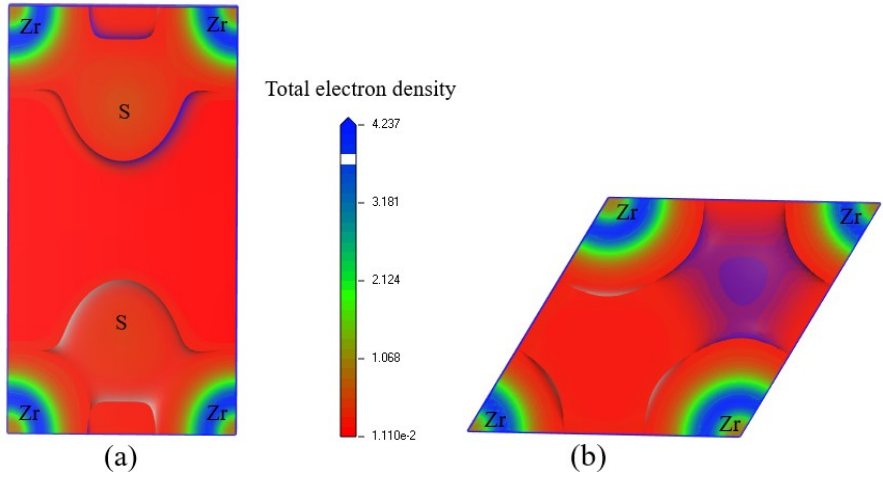

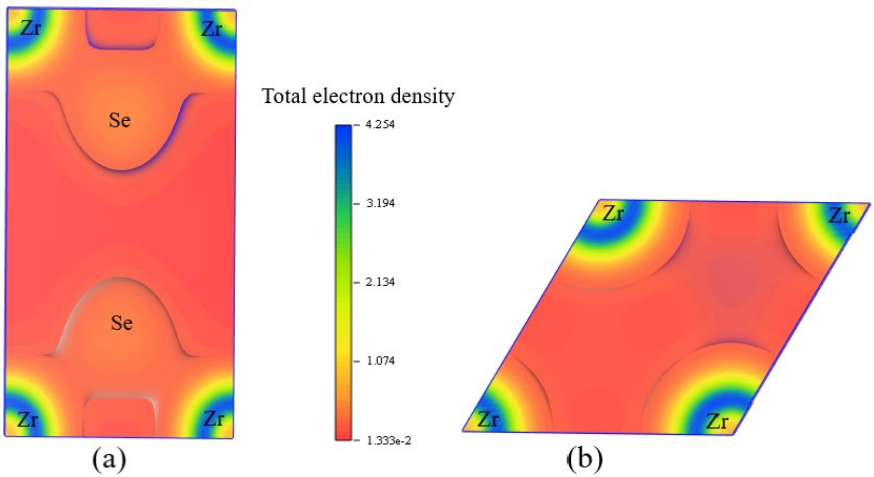

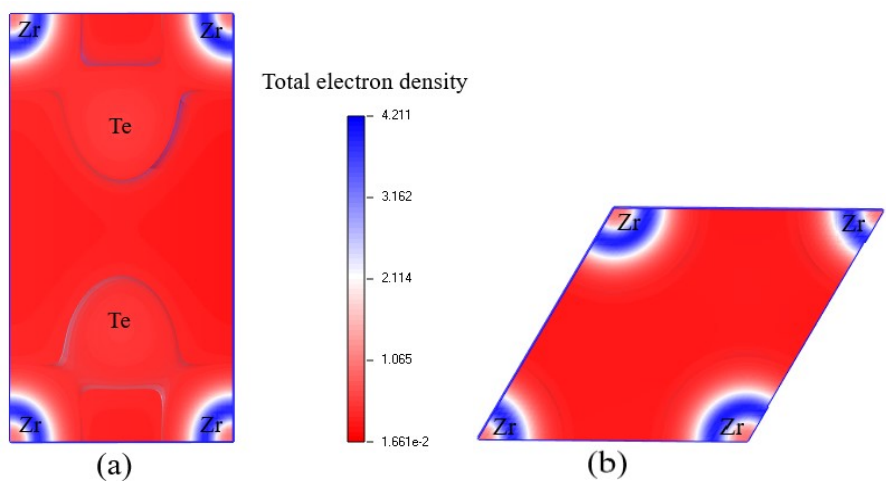

**Figure 5.** Charge density distribution map of ZrX$_2$ (X = S, Se, Te) in (a) (100) and (b) (001) planes.

From the above charge density distribution maps we see that the charge distribution around the X atomic species gets significantly distorted of spherical shape due to the charge distribution of the surrounding Zr atoms. This is an indication of covalent bonding. Since charge on X atoms is higher than those of the Zr atoms, there is also ionic contribution. Figs. 5b for ZrS$_2$ and ZrSe$_2$ show directional bonding between the Zr atoms where clear signs of charge accumulation between those are seen. This is indicative of the formation of covalent bonding between these atoms. It is interesting to note that such charge accumulation is absent for the metallic ZrTe$_2$ compound. Almost uniform charge density between the Zr atoms in ZrTe$_2$ implies that bonding between these atoms is metallic in nature. This perhaps explains why the hardness of ZrTe$_2$ is much lower than that for ZrS$_2$ or ZrSe$_2$.

**(b) Bonding properties**

We have studied both Mulliken population analysis (MPA) [88] and Hirshfeld population analysis (HPA) [89] to explore the bonding nature of ZrX$_2$ (X = S, Se, Te) compounds further. The results are disclosed in Tables 7 and 8. According to MPA, in ZrS$_2$ Zr atom gives up 0.29 charge to each S atom and in ZrTe$_2$ two Te atoms transfer 0.49e charge to Zr. On the other hand, no charge transfer occurs between Zr and Se atom in ZrSe$_2$. The transfer of electrons between different atoms in the compounds is due to the partial presence of ionic bonding. The difference between formal ionic charge and calculated Mulliken charge is called the effective valence charge (EVC). Non-zero EVC in all the three compounds is an indication of covalent bonding among the atoms in all these compounds (Table 7). On the other hand, according to HPA, in ZrS$_2$ compound Zr atom 0.14e electronic charge to each S atom. This result is consistent with that of MPA. In ZrSe$_2$, Zr atom transfers 0.10e charge to each Se atom. But in ZrTe$_2$, very small amount of electronic charge is transferred to Te from Zr, which is completely opposite to MPA. Overall, the HPA results are more reliable because it is free from basis set dependence. In contrast,

Mulliken population analysis may lead to large changes of the computed atomic charges for small variations in the underlying basis sets and can overestimate the covalent character of a bond.

Table 7. Charge spilling parameter (%), orbital charges (electron), atomic Mulliken charge (electron), Hirshfeld charge (electron), and EVC (electron) of $ZrX_2$ (X = S, Se, Te).

| Compound | Atoms | Charge spilling (%) | s | p | d | Total | Mulliken charge | Hirshfeld charge | EVC |
|---|---|---|---|---|---|---|---|---|---|
| $ZrS_2$ | S | 1.13 | 1.85 | 4.44 | 0.0 | 6.27 | -0.29 | -0.14 | 1.71 |
| | Zr | | 2.43 | 6.48 | 2.52 | 11.42 | 0.58 | 0.28 | 3.42 |
| $ZrSe_2$ | Se | 0.78 | 1.70 | 4.30 | 0.0 | 6.0 | 0.0 | -0.10 | 2.0 |
| | Zr | | 2.58 | 6.89 | 2.59 | 12.0 | 0.0 | 0.20 | 4.0 |
| $ZrTe_2$ | Te | 0.85 | 1.67 | 4.08 | 0.0 | 5.75 | 0.25 | -0.03 | 1.75 |
| | Zr | | 2.66 | 6.99 | 2.84 | 12.49 | -0.49 | 0.07 | 3.51 |

From Table 8, it is seen that the value of overlap population are small and positive for each material. The small positive value indicates there is weak interaction between the atoms and they have bonding nature. Generally, shorter bond length represents stronger bonding and consequently higher bond strengths. Among the $ZrX_2$ (X = S, Se, Te) TMDCs, the bond lengths of Zr-S and Zr-Se are quite close. For Zr-Te, the bond lengths are higher and the bond populations are lower. This implies that the bonding strengths in $ZrTe_2$ are significantly lower. These results are completely in accord with the hardness values of $ZrX_2$ (X = S, Se, Te) obtained in the preceding section.

Table 8. Calculated bond overlap population and bond lengths (Å) for $ZrX_2$ (X = S, Se, Te) TMDCs.

| Compound | Bond | Population | Length |
|---|---|---|---|
| $ZrS_2$ | S2 -- Zr1 | 1.30 | 2.551 |
| | S1 -- Zr1 | 1.30 | 2.551 |
| $ZrSe_2$ | Se2 -- Zr1 | 0.87 | 2.663 |
| | Se1 -- Zr1 | 0.87 | 2.663 |
| $ZrTe_2$ | Zr1 -- Te1 | 0.55 | 2.892 |
| | Zr1 -- Te2 | 0.55 | 2.892 |

## V. Optical properties

The understanding of energy/frequency dependent optical parameters is essential to predict how a material will respond when electromagnetic radiation is incident on it. In order to investigate possible optoelectronic applications of a compound, knowledge regarding the response of the compound to infrared, visible and ultraviolet spectra is important. Various frequency dependent optical constants, namely, dielectric function $\varepsilon(\omega)$, refractive index $n(\omega)$, optical conductivity $\sigma(\omega)$, reflectivity $R(\omega)$, absorption coefficient $\alpha(\omega)$ and energy loss function $L(\omega)$ (where $\omega = 2\pi f$ is the angular frequency) are calculated in this section to explore the response of $ZrX_2$ (X = S, Se, Te) to incident photons. The optical parameters of the compounds are depicted in Figs. 6, 7 and 8 for incident energy up to 25 eV and the electric field polarizations along [100], [010], and [001] directions.

Figs. 6(a), 7(a) and 8(a) represent the absorption coefficient $\alpha(\omega)$ of $ZrX_2$ (X = S, Se, Te). The absorption coefficients start at around 0.75 eV and 0.20 eV for $ZrS_2$ and $ZrSe_2$ respectively, which confirms the semiconducting nature of these compounds. The onset energies also agree very well with the band gap values obtained from the band structure calculations. In $ZrTe_2$, $\sigma(\omega)$ starts from 0 eV, which is an indication of its metallic nature. All the compounds in this study exhibit a peak in the absorption spectrum in the ultraviolet region which falls down in the higher energy region for all the polarizations of the electric field. The absorption spectra suggest that these compounds can absorb ultraviolet radiation quite efficiently. The absorption spectra originate mainly from the photon induced electronic transitions between the S-3p, Se-4p, Te-5p, and Zr-4d electronic states in the valence and conduction bands for $ZrS_2$, $ZrSe_2$ and $ZrTe_2$, respectively. Figures 6(b), 7(b) and 8(b) show the real and imaginary parts of optical conductivity $\sigma(\omega)$. For $ZrS_2$ and $ZrSe_2$ real part of $\sigma(\omega)$ start at around 0.75 eV and 0.20 eV respectively, which is another indication of their semiconducting nature. On the other hand, optical conductivity is finite at zero energy in $ZrTe_2$ indicating again its metallic character. The real and imaginary parts of dielectric constants are illustrated in Figures 6(c), 7(c) and 8(c). The real part of the dielectric constant is related to the electrical polarization of the material, while the imaginary part is linked with dielectric loss. The real part crosses zero from below at ~18.5 eV for $ZrS_2$ and $ZrSe_2$ and at ~20 eV for $ZrTe_2$. After that this part approaches unity. On the other hand, the imaginary part falls sharply and flattens to a low value at the same energy.

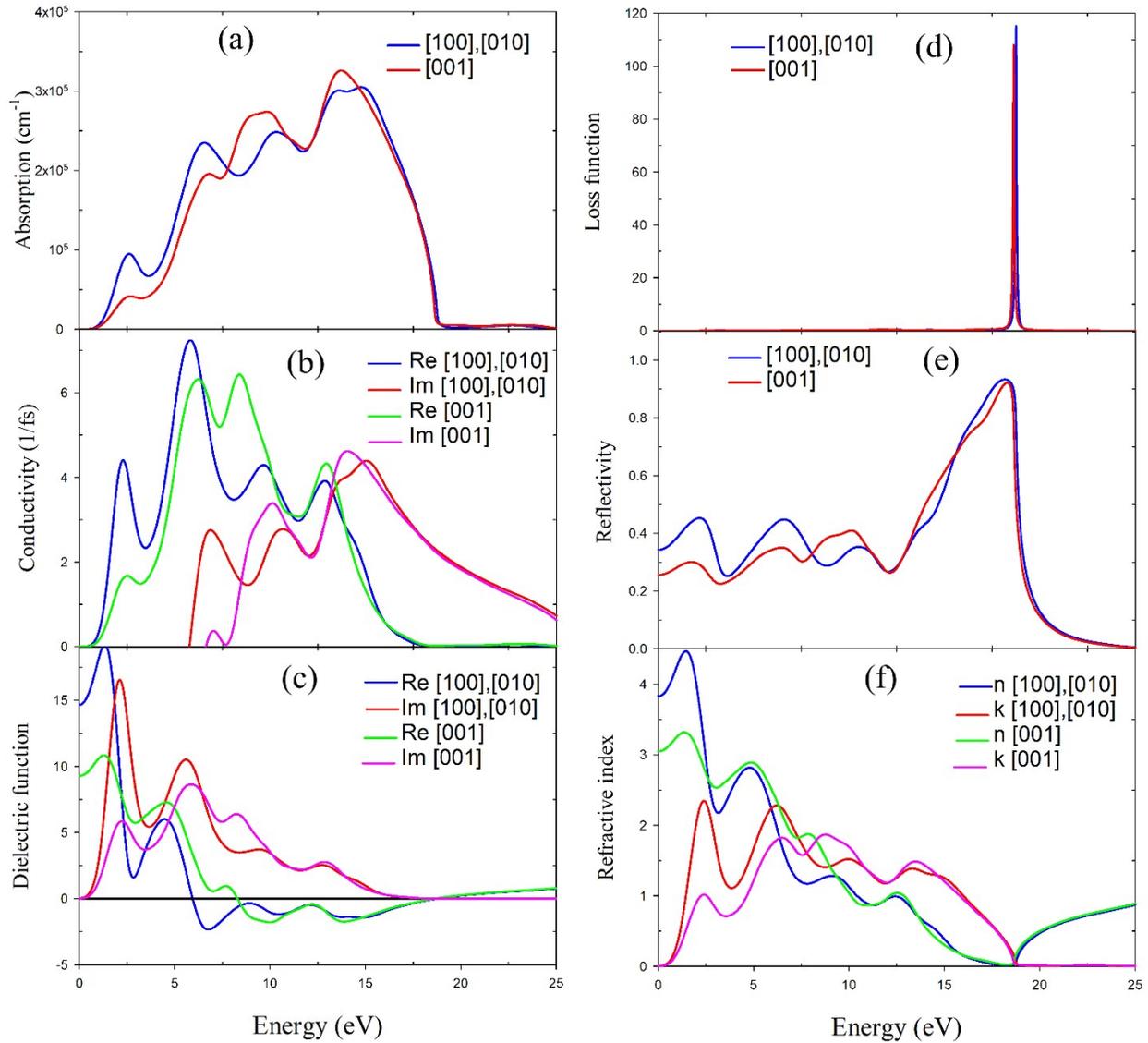

**Figure 6.** The frequency dependent (a) absorption coefficient (b) optical conductivity (c) dielectric function (d) loss function (e) reflectivity and (f) refractive index of $ZrS_2$ with electric field polarization vectors along [100], [010], and [001] directions.

The loss function $L(\omega)$ of $ZrX_2$ (X = S, Se, Te) are depicted in Figures6(d), 7(d) and 7(d). The loss peaks are found at ~18.5 eV for $ZrS_2$ and $ZrSe_2$ and at ~20 eV for $ZrTe_2$. These peaks mark the characteristic plasmon energy for corresponding materials. The plasma oscillations due to collective motions of the charge carriers are induced at these particular energies. It is worth noticing that the plasma energies coincide with the sharp falls in the absorption coefficient and reflectivity. This implies that the compounds under investigation are expected to behave transparently for photons with energies greater than the plasma energy and the optical properties will show behaviors appropriate for insulating systems.

The reflectivity profile for ZrX$_2$ (X = S, Se, Te) compounds are shown in Figures 6(e), 7(e) and 8(e). The reflectivity spectra fall sharply at plasma frequency in all three compounds.

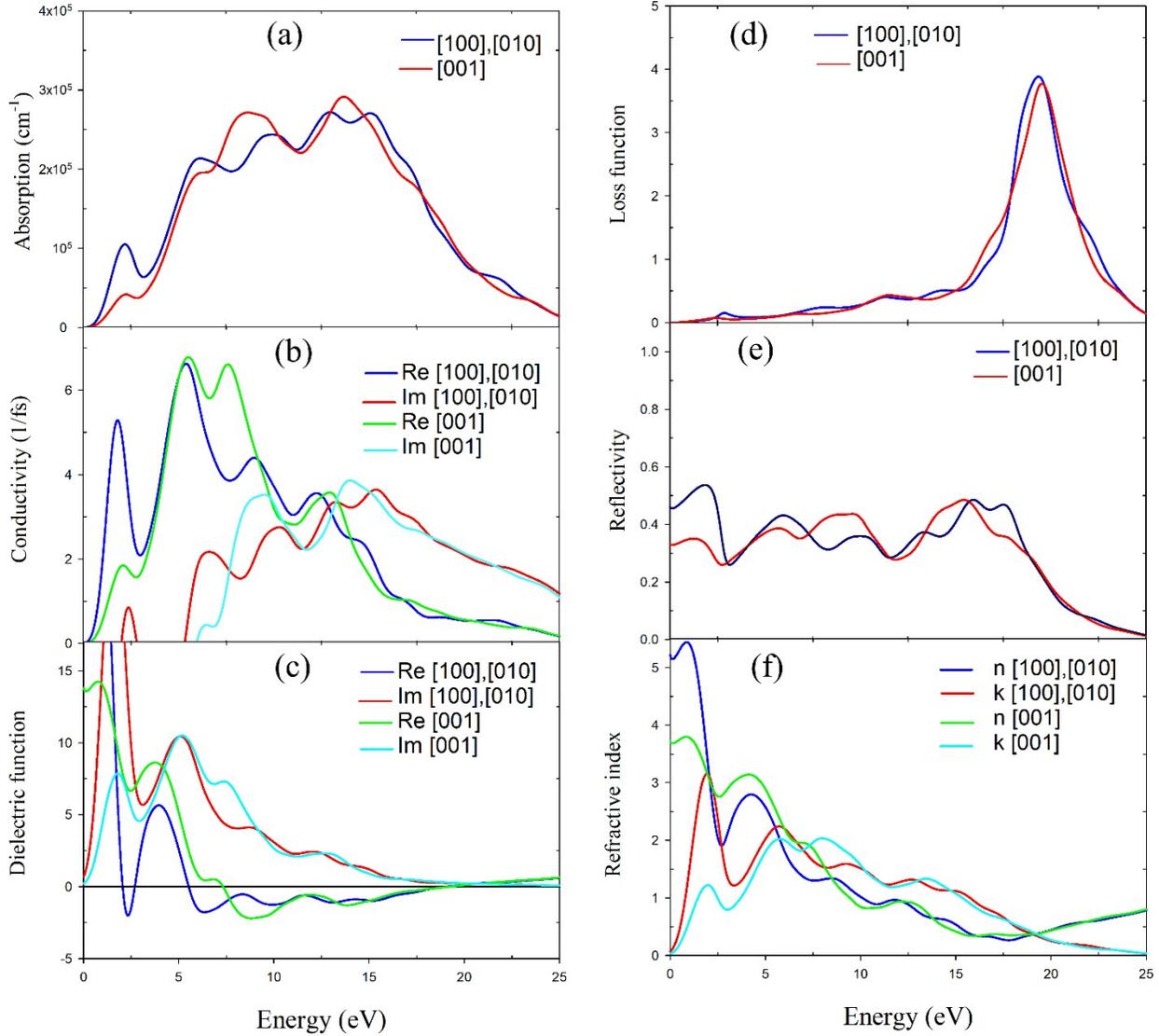

**Figure 7.** The frequency dependent (a) absorption coefficient (b) optical conductivity (c) dielectric function (d) loss function (e) reflectivity and (f) refractive index of ZrSe$_2$ with electric field polarization vectors along [100], [010], and [001] directions.

Reflectivity of ZrS$_2$ is low in the visible and near-ultraviolet region but R($\omega$) rises sharply in the mid-ultraviolet energies and approaches 95% at ~18 eV. Therefore, this compound has the potential to be used as an efficient ultraviolet reflector in the mid-ultraviolet region. R($\omega$) is relatively low over a very wide energy band and displays almost non-selective character. In ZrTe$_2$, R($\omega$) remains over 50% in the energy range from 0 eV to ~20 eV. So, this material has wide band high reflectivity and can be employed as a reflector to reduce solar heating. The frequency dependent real and imaginary parts of refractive index are represented in Figs. 6(f), 7(f) and 8(f). The phase velocity of electromagnetic wave in the material is determined by the real part of refractive index, whereas the attenuation of electromagnetic radiation inside the material is measured by the imaginary part, often referred to as extinction coefficient. The real part of refractive index has high value at low energies (0-5 eV) for both $ZrS_2$ and $ZrSe_2$.

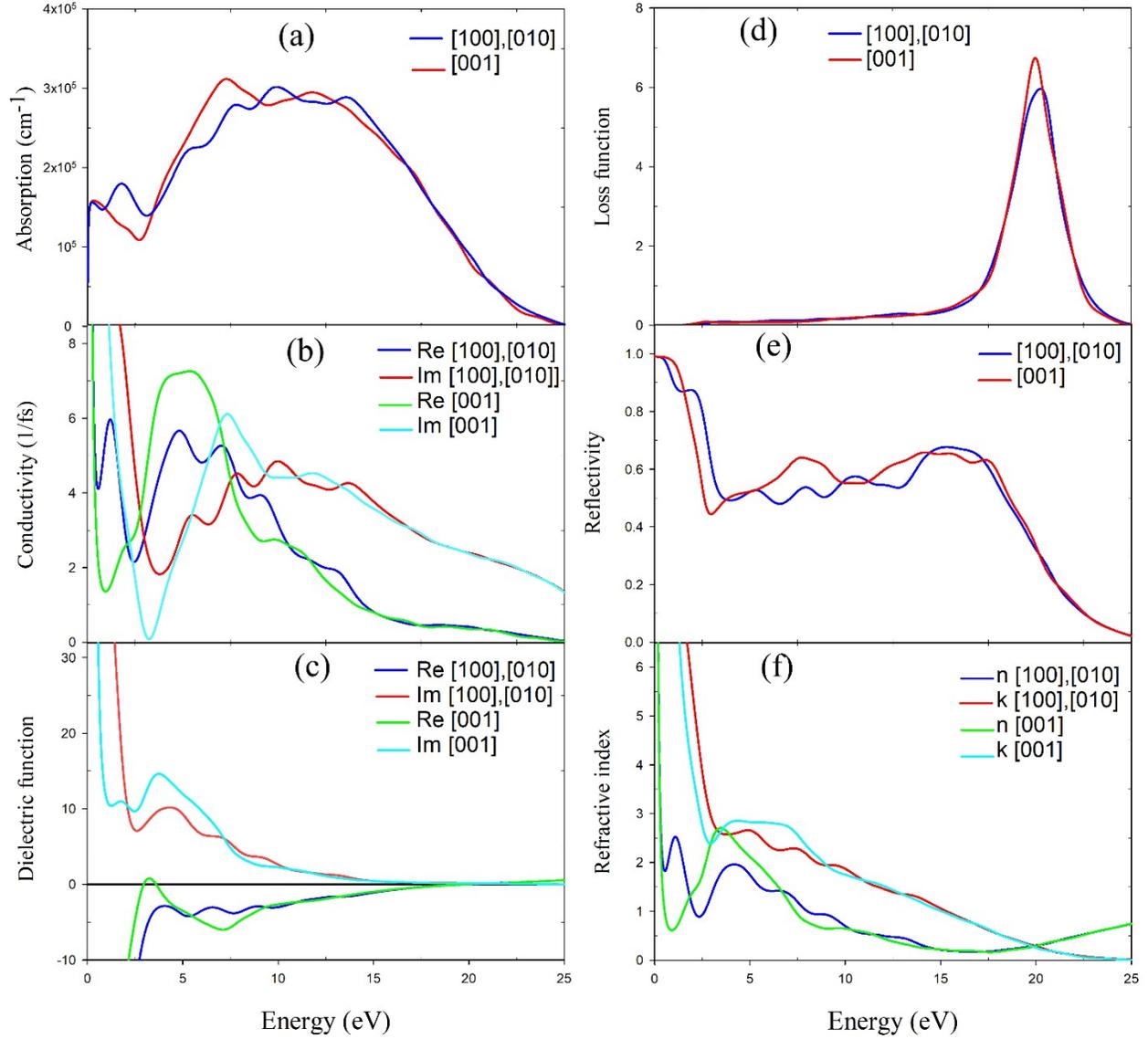

**Figure 8.** The frequency dependent (a) absorption coefficient (b) optical conductivity (c) dielectric function (d) loss function (e) reflectivity and (f) refractive index of $ZrTe_2$ with electric field polarization vectors along [100], [010], and [001] directions.

Therefore, the materials under study have desired optical characteristics for light emitting and optoelectronic display devices. All the optical parameters of $ZrX_2$ (X = S, Se, Te) show moderate anisotropy with respect to the polarization direction of the electric field. Compared to other optical parameters, optical anisotropy in the refractive index is the most prominent.

## VI. Thermo-physical properties

The Debye temperature, $\theta_D$, corresponds to the highest frequency normal mode of vibration in a crystal. This temperature depends on the crystal stiffness and constituent atomic masses of the compound. $\theta_D$ is a fundamental thermo-physical parameter of solids, which is related to many physical properties such as lattice thermal conductivity, phonon specific heat, melting temperature, vacancy formation energy, bonding strength among the atoms within the crystal etc. At low temperatures only acoustic modes are responsible for vibrational excitations. Thus, at low temperature, $\theta_D$ calculated from elastic constants and specific heat are identical. In this study, $\theta_D$ is calculated from its proportionality to the mean sound velocity inside the crystal as [90],

$$\theta_D = \frac{h}{k_B}\left[\left(\frac{3n}{4\pi}\right)\frac{N_A \rho}{M}\right]^{\frac{1}{3}} v_m \tag{22}$$

where h is the Planck's constant, $k_B$ is the Boltzmann's constant, n denotes number of atoms within the unit cell, M is molar mass, $\rho$ is density, $N_A$ is Avogadro's number and $v_m$ denotes mean sound velocity. $v_m$ can be determined from bulk (B) and shear (G) modulus through longitudinal ($v_l$) and transverse ($v_t$) sound velocities as follows,

$$v_m = \left[\frac{1}{3}\left(\frac{2}{v_t^3} + \frac{1}{v_l^3}\right)\right]^{-\frac{1}{3}} \tag{23}$$

where,

$$v_t = \sqrt{\frac{G}{\rho}} \tag{24}$$

$$v_l = \sqrt{\frac{3B + 4G}{3\rho}} \tag{25}$$

The calculated Debye temperature, $\theta_D$ with $v_l$, $v_t$ and $v_m$ are enlisted in Table 9. It is well known that, a higher Debye temperature denotes a higher phonon thermal conductivity. Fine *et al.* [91] proposed an empirical formula to determine the melting temperature of solids by using the elastic constants as follows:

$$T_m = 354 + 1.5(2C_{11} + C_{33}) \tag{26}$$

With the help of thermal conductivity the behavior of atoms inside a crystal can be understood when the crystal is heated or cooled. The hypothetical lowest value of inherent thermal conductivity is defined as the minimum thermal conductivity. This parameter is determined by the equation as follows [92–94]:

$$k_{min} = k_B v_m \left(\frac{nN_A\rho}{M}\right)^{\frac{2}{3}} \quad (27)$$

The calculated values of $T_m$ and $k_{min}$ are also given in Table 9. From Table 4 and 9, it is seen that among the TMDCs under study, the Debye temperature is the largest for the compound which has highest melting temperature and highest hardness. This is expected because larger Debye temperature indicates stronger interatomic bonding that leads to higher melting temperature and elevated mechanical strength. Overall, the calculated Debye temperatures are low for all the $ZrX_2$ (X = S, Se, Te) compounds indicating the soft nature of these solids. The estimated $\theta_D$ of $ZrS_2$ shows very good agreement with previously obtained result [95]. Compared to many other binaries the melting temperatures of $ZrX_2$ (X = S, Se, Te) are quite low. The minimum phonon thermal conductivities are also low. Clear correspondence among $\theta_D$, $T_m$, and $k_{min}$ are observed in $ZrX_2$ (X = S, Se, Te).

**Table 9.** Calculated mass density ($\rho$ in gm cm$^{-3}$), longitudinal, transverse and sound velocities ($v_l$, $v_t$ and $v_m$ in km sec$^{-1}$), and Debye temperature ($\theta_D$ in K), melting temperature ($T_m$ in K) and minimum thermal conductivity ($k_{min}$ in Wm$^{-1}$K$^{-1}$) of $ZrX_2$ (X = S, Se, Te) compounds.

| Compounds | $\rho$ | $v_l$ | $v_t$ | $v_m$ | $\theta_D$ | $T_m$ | $k_{min}$ | Ref. |
|---|---|---|---|---|---|---|---|---|
| $ZrS_2$ | 3.85 | 3.822 | 2.369 | 2.612 | 276.44 | 779.98 | 0.0454 | This |
|  | - | - | - | 3.996 | 290.87 | - | - | [92] |
| $ZrSe_2$ | 5.72 | 3.437 | 2.20 | 2.417 | 249.25 | 750.20 | 0.0399 | This |
| $ZrTe_2$ | 6.63 | 2.546 | 1.485 | 1.647 | 159.98 | 605.39 | 0.0241 | This |

## 4. Discussion and conclusions

Structural, elastic, electronic, boding, optical and some thermal properties of TMDCs, $ZrX_2$ (X = S, Se, Te) compounds have been studied via ab-initio technique. Elastic anisotropy, atomic bonding character, hardness and machinability of $ZrX_2$ (X = S, Se, Te) are investigated for the first time. The band gaps of $ZrS_2$ and $ZrSe_2$ show systematic variation with c-axis lattice parameter and eventually vanishes (valence band overlaps with the conduction band) for $ZrTe_2$ which has a significantly larger c-axis lattice parameter compared to the other two. The change in the crystal volume closely follows the changes in atomic radii as one progressively moves from S to Te. The structural results show close agreement with previous experimental and theoretical results where available. The machinability index of $ZrX_2$ (X = S, Se, Te) are quite high, particularly so of $ZrS_2$. All three TMDCs are mechanically anisotropic with mixed bonding character. Compared to $ZrS_2$ and $ZrSe_2$, $ZrTe_2$ is significantly softer. The optical spectra are studied for the first time and also disclose several interesting features. All three TMDCs are good absorbers of ultraviolet radiation. Reflectivity of $ZrS_2$ is low in the visible and near-ultraviolet region but it increases sharply for higher photon energies and approaches 95% at ~18 eV. $R(\omega)$ spectrum of $ZrTe_2$, on the other hand, is non-selective and stays above 50% over a wide range of energies encompassing infrared to ultraviolet. This compound has the potential to be used as an efficient solar reflector. Furthermore, the refractive indices of $ZrS_2$, $ZrSe_2$ and $ZrTe_2$ in the visible range are high. Optical spectra show moderate anisotropy with respect to the polarization of the electric field of the incident light. Debye temperatures of the compounds under investigation are low, particularly of $ZrTe_2$. Very low $\theta_D$ of $ZrTe_2$ is a consequence of significantly softer lattice of this compound compared to the other two. Low values of estimated melting temperatures and minimum phonon thermal conductivity correspond very well to the calculated values of hardness and various calculated elastic moduli.

To summarize, we have presented a comprehensive study of a number of important physical properties of TMDCs $ZrX_2$ (X = S, Se, Te) from the point of view of their possible applications. New results should be treated as predictions and should serve as references for future studies.

## Acknowledgements

S.H.N. and R.S.I. acknowledge the research grant (1151/5/52/RU/Science-07/19-20) from the Faculty of Science, University of Rajshahi, Bangladesh, which supported the computational section of this work.

## Data availability

The data sets generated and/or analyzed in this study are available from the corresponding author on reasonable request.

**Author Contributions**

**Additional Information**

**Competing Interests**

The authors declare no competing interests.